\documentclass{llncs}

\usepackage{graphicx}
\usepackage{mathptmx}
\usepackage{tikz-qtree}
\usepackage{pgfplots}
\usepackage{filecontents}
\usepackage{csvsimple}
\usepackage{amsfonts,amsmath}
\usepackage{url}
\usepackage{verbatim}
\usepackage{color}

\definecolor{lgray}{gray}{0.95}
\definecolor{lblue}{rgb}{0.90,0.90,1.00}
\definecolor{lyellow}{rgb}{1.00,1.00,0.70}

\usepackage{listings}
\newtheorem{ex}{Example}

\newenvironment{codex}{\small\verbatim}{\endverbatim\normalsize}

\lstnewenvironment{code}
    {\lstset{}%
      \csname lst@SetFirstLabel\endcsname}
    {\csname lst@SaveFirstLabel\endcsname}
\newcommand{\BI}[0]{\begin{itemize}}
\newcommand{\EI}[0]{\end{itemize}}
\newcommand{\I}[0]{\item}
\newcommand{\BE}[0]{\begin{enumerate}}
\newcommand{\EE}[0]{\end{enumerate}}

\newcommand{\BX}[0]{\begin{ex}}
\newcommand{\EX}[0]{\end{ex}}
\newcommand{\BV}[0]{\begin{verbatim}}
\newcommand{\EV}[0]{\end{verbatim}}

\newcommand{\BF}[0]{\begin{filecontents*}{data.csv}}

\newcommand{\BQ}[0]{\color{blue}\begin{quote}}
\newcommand{\EQ}[0]{\end{quote}\color{black}}

\def \bscale1 {0.25}
\def \bscale {0.25}

\begin{document}

\title{Formula Transformers and Combinatorial Test 
Generators for Propositional Intuitionistic Theorem Provers}

\author{Paul Tarau}

\institute{
   {Department of Computer Science and Engineering}\\
   {University of North Texas}\\
   {\em paul.tarau@unt.edu}
}

\maketitle

\begin{abstract}

We  develop combinatorial test generation algorithms for  progressively more powerful theorem provers, covering formula languages ranging from the implicational fragment of intuitionistic logic to full intuitionistic propositional logic.
Our  algorithms support  exhaustive and random generators for  formulas of these logics. To provide known-to-be-provable formulas, via the Curry-Howard formulas-as-types  correspondence, we use generators for typable lambda terms and combinator expressions. 
Besides generators for several classes of formulas, 
we design algorithms that restrict formula generation
to canonical representatives among equiprovable formulas and introduce
program transformations that reduce formulas to equivalent formulas of a
simpler structure. The same transformations, when applied in reverse, 
create harder formulas that can catch soundness or incompleteness bugs.
   To test the effectiveness of the testing framework itself, we describe use cases for deriving lightweight theorem provers for several of these logics and for finding bugs in known theorem provers.
Our Prolog implementation available at:\\
{\bf \url{https://github.com/ptarau/TypesAndProofs}} and a subset of formula generators and theorem provers, implemented in 
Python is available at:\\
 {\bf \url{https://github.com/ptarau/PythonProvers}}.


\end{abstract}

\section{Introduction}

Theorem provers have been used in the last half-century not just to solve interesting mathematical problems but also in important practical
software and hardware verification projects, ranging from 
nuclear reactor controllers and space-ship components 
to pacemakers and floating point units.

Correctness 
and performance of theorem provers are usually being tested
using comprehensive repositories of ``human-made''
problems, such as the TPTP library\footnote{\url{http://tptp.cs.miami.edu/~tptp/}}
for classical logic or the ILTP library\footnote{\url{http://www.iltp.de/}} for
intuitionistic logic. Similarly, extensive online 
benchmarks help evaluate  SAT, SMT and ASP solvers. 
Besides their chance to spot out correctness and scalability issues,
``human-made'' test libraries, often derived from
interesting mathematical problems, can measure the  ability of
theorem provers and solver engines to work well on specific 
problem domains.

However, as automation of testing is gaining significant traction in
both software and hardware validation, it is natural
to think about adopting automated testing techniques for theorem provers, 
which are, after all, software artifacts. 
Even if some of the   ``human-made'' test sets have accumulated over the years
hundreds and often thousands of problem instances, truly ``adversarial''
computer generated correctness and scalability tests can spot out soundness, completeness
or termination issues overlooked by implementors of the intricate, heuristic-driven code of
today's theorem provers.
At the same time, validation via a trusted,
``gold-standard'' prover, producing the same results 
on the same test set,
can be used to propagate incrementally
correctness of provers from simple formally validated versions
to more sophisticated versions implementing complex heuristics.

Designing the algorithms that generate tests for theorem provers
is facilitated by the regular structure of their input formulas.
Such tests can be based on exhaustive small formula generators
as well as   random large formula generators.
Besides comparison with lightweight versions of the provers,
for which correctness is formally provable (possibly via
proof assistants like Coq \cite{Coq:manual}
or Agda \cite{agda}),
known isomorphisms between formula languages and
computational mechanisms like  typed lambda calculi
offer opportunities for transferring properties across
``bridges'' like the Curry-Howard {\em formulas-as-types}
correspondence \cite{howard:formulaeastypes:hbc:80,wadler15}, that 
 ensures that the inferred type of a lambda term
is a tautology in intuitionistic logic.

While the problem of finding a lambda term that has a given simple type
(called the {\em inhabitation problem})
is PSPACE-complete  \cite{statman79}, 
efficient algorithms have been
known for a long time for inferring the simple type of a
lambda expression, when it exists \cite{hindley2008lambda}. 
The key step in the ``inner loop'' of
this process is {\em unification with occurs-check}
\cite{robinson:machine:jacm:65}, for which
today's Prolog systems offer highly efficient implementations.

The symbiosis between Automated Theorem Proving and  and Logic Programming 
has been observed in the evolution of both research fields as early as
in \cite{wosSym}.
With sound unification and backtracking efficiently implemented in today's
logic programming languages (e.g., Prolog, Curry, Picat), 
one can take advantage of the natural synergy that exists in these
languages with features like
Definite Clause Grammars (DCGs), to provide together
an ideal playground for exploring  combinatorial
properties of  typed lambda terms \cite{arxiv_play15} and 
corresponding formula languages, 
essential for their  applications to  generation of very large terms and 
valid formulas. They also provide an ideal framework for transliterating
sequent calculus rules into executable code.

We have built our Prolog-based open-source testing framework
covering combinatorial test generators and several lightweight theorem provers for
propositional intuitionistic and classical logics.
Beside several of our own and 3-rd party provers, the github site\footnote{
\url{https://github.com/ptarau/TypesAndProofs}
}
contains test generators and formula readers converting the ``human-made'' tests at \url{http://www.iltp.de} to Prolog and Python form.
Part of our testing framework focusing on the implicational fragment of intuitionistic logic is described in  \cite{padl19} where examples of step-by step, test-driven derivations of provers are given. 
More recently, using these tests on the full intuitionistic propositional calculus has revealed interesting examples of Byzantine failures occurring with some of the 3-rd party provers we tested. For instance, increasing the standard 600 second
timeout  has revealed  cases of non-termination leading to stack overflows and
unexpected space complexity resulting in heap overflows, when given  a very large RAM (e.g., 64GB or 96GB).

The two main generator families implemented in our  testing framework are exhaustive formula generators and random formula generators. Exhaustive formula generators enumerate all formulas of a given size and thus are useful for finding minimal failure instances for a given incorrect prover. Random formulas, especially if generated as known-to-be-provable, besides pointing out soundness bugs, are also relevant as scalability tests, catching unexpected space or computation time explosion.

At the same time our testing framework contains several representation transformers that convert classes of formulas to equivalent or equiprovable\footnote{Two formulas are called equiprovable if, finding a proof for one entails the existence of a proof for the other. In particular, logically equivalent formulas are equiprovable.} formulas. 

Through a series of use cases, we exhibit provers obtained via test-driven refinements
and discuss their improvements in performance and reduced space complexity.

We summarize here the main contributions of this paper:
\BI
\I new combinatorial test generation algorithms for (sub-)formula languages of intuitionistic propositional logic
\I restriction mechanisms limiting formula generators to one representative per class of equiprovable formulas
\I transformers from disjunction-free formulas to a Nested Horn Clause form reducing space complexity from exponential to $O(n~log(n))$
\I several lightweight theorem provers obtained as a result of test-driven refinements using our formula generators
\I use cases showing effectiveness of our framework in finding bugs in theorem provers
\EI

The rest of the paper is organized as follows.

Section \ref{exh} describes  exhaustive generation algorithms for formulas
of given (small) size, covering formulas known to be tautologies as well
as arbitrary formulas.
Section \ref{rand} describes algorithms generating random formulas of the same two categories.
Section \ref{trans} introduces mechanisms restricting formula generators to canonical
representatives of equivalence classes as well as transformations to equivalent, structurally
simpler formulas.
Section \ref{use} describes test-driven refinements of provers derived from sound and complete calculi result in significant performance or space complexity improvements.
Section \ref{tests} overviews our combinatorial testing framework.
Section \ref{bugs} shows the effectiveness of our framework in finding bugs in theorem provers.
Section \ref{rel} discusses related work and
section \ref{conc} concludes the paper.

\section{Exhaustive Formula Generation Algorithms}\label{exh}

An advantage of exhaustive testing with all formulas of a given size is that it implicitly ensures full coverage: no path is missed simply because there are no paths left unexplored.

\subsubsection*{Notations and Assumptions}

As we will use {\bf Prolog} as our meta-language, our notations will be derived 
as much as possible from its syntax (including token types and operator
definitions). Thus, variables will be denoted with uppercase letters
and, as programmer's conventions final {\tt s} letters indicate
a plurality of items (e.g., when referring to the content
of $\Gamma$ contexts). 
We assume that the reader is familiar with basic Prolog programming,
including, besides the pure Horn clause subset, well-known builtin predicates
like {\tt memberchk/2} and {\tt select/3}, elements of higher order
programming (e.g., {\tt call/N}), 
and occasional use of CUT and {\tt if-then-else} constructs.

Lambda terms are built using the function symbols {\bf a/2}=application, {\bf l/2}=lambda binder, with a logic variable as first argument and expression as second, 
as well as {\em logic variables} representing the variables of the terms.

Type expressions (also seen as implicational formulas) are built as binary trees
with the function symbol {\verb~->/2~ and {\em logic variables at their leaves}.

\BX{The {\bf S} combinator (left) and its type (right, with integers as leaves):}\\
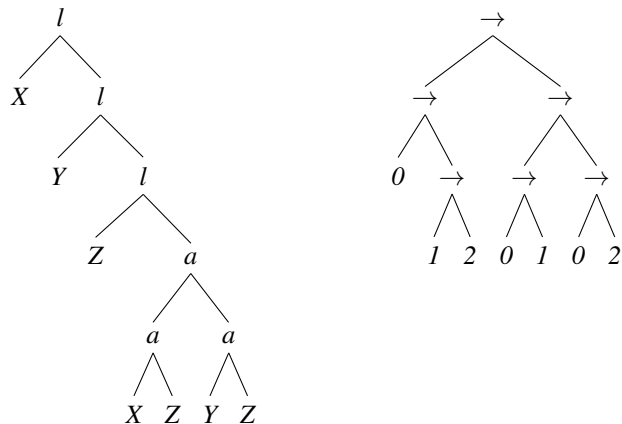
\begin{center}
\Tree [.l [.X ] [.l [.Y ] [.l [.Z ] [.a [.a [.X ] [.Z ]  ] [.a [.Y ] [.Z ]  ]  ]  ]  ]  ]
~~~~~~~~~~~~~~~~~
\Tree [.$\rightarrow$ [.$\rightarrow$ [.0 ] [.$\rightarrow$ [.1 ] [.2 ]  ]  ] [.$\rightarrow$ [.$\rightarrow$ [.0 ] [.1 ]  ] [.$\rightarrow$ [.0 ] [.2 ]  ]  ]  ]
\end{center}
\EX

\subsection{The Language of Implicational Formulas} \label{ltypes}

As a result of the Curry-Howard correspondence,
the language of types is isomorphic with that
of {\em the implicational fragment of intuitionistic propositional logic}, 
with binary trees having variables
at leaf positions and the implication operator (``\verb~->~'')
at internal nodes. We will rely on the right associativity
of this operator in Prolog, that matches the standard
notation in type theory.

The predicate {\tt type\_skel/3} generates all binary trees
with given number of internal nodes and it labels their leaves
with unique logic variables. It also collects the leaf variables
to a list returned as its third argument.
\begin{code}
type_skel(N,T,Vs):-type_skel(T,Vs,[],N,0).

type_skel(V,[V|Vs],Vs)-->[].
type_skel((X->Y),Vs1,Vs3)-->pred,type_skel(X,Vs1,Vs2),type_skel(Y,Vs2,Vs3).
\end{code}
Type skeletons are counted by the Catalan numbers
(sequence {\tt A000108} OEIS in \cite{intseq}).
\BX
All type skeletons for N=3.
\begin{codex}
?- type_skel(3,T,_).
T =  (A->B->C->D) ; T =  (A-> (B->C)->D) ; T =  ((A->B)->C->D) ;
T =  ((A->B->C)->D) ; T =  (((A->B)->C)->D) .
\end{codex}
\EX
The mechanism is extended to use additional constructors
from the set \verb|{~,&,v,<->}| as internal nodes of the generated trees,
to cover the language of full intuitionistic propositional calculus\
\footnote{at \url{https://github.com/ptarau/TypesAndProofs/blob/master/allFormulas.pro}}
.

The next step  toward generating the set of all type formulas
is observing that logic variables define equivalence
classes that  correspond to partitions
of the set of variables, simply by
selectively unifying them.

The predicate {\tt mpart\_of/2}
takes a list of distinct logic variables
and generates partitions-as-equivalence-relations
by unifying them ``nondeterministically''.
It also collects the unique variables defining
the equivalence classes, as a list given by its second argument.
\begin{code}
mpart_of([],[]).
mpart_of([U|Xs],[U|Us]):-mcomplement_of(U,Xs,Rs),mpart_of(Rs,Us).
\end{code}

To implement a set-partition generator,
we  split a set repeatedly in subset+complement
pairs with help from the predicate {\tt mcomplement\_of/2}.
\begin{code}
mcomplement_of(_,[],[]).
mcomplement_of(U,[X|Xs],NewZs):-
  mcomplement_of(U,Xs,Zs),
  mplace_element(U,X,Zs,NewZs).

mplace_element(U,U,Zs,Zs).
mplace_element(_,X,Zs,[X|Zs]).
\end{code}
To generate all set partitions of a set of variables
of a given size, we build a list
of fresh variables with 
Prolog's {built-in predicate {\tt length/2}
and constrain {\tt mpart\_of/2} to use them as the set to be partitioned.
\begin{code}
partitions(N,Ps):-length(Ps,N),mpart_of(Ps,_).
\end{code}
The counts of the resulting set-partitions (Bell numbers)
corresponds to the entry {\tt A000110} 
in \cite{intseq}.

\BX
Set partitions of size 3 expressed as variable equalities.
\begin{codex}
?- partitions(3,P).
P = [A, A, A]; P = [A, B, A]; P = [A, A, B]; P = [A, B, B]; P = [A, B, C].
\end{codex}
\EX

Hence, we can define the language of formulas in implicational intuitionistic propositional
logic, among which tautologies will correspond to
simple types, as being generated by the predicate {\tt maybe\_type/3}.
\begin{code}
maybe_type(L,T,Us):-type_skel(L,T,Vs),mpart_of(Vs,Us).
\end{code}

\BX
Well-formed  formulas of the implicational fragment of intuitionistic propositional logic (possibly types) of size 2.

\begin{codex}
?- maybe_type(2,T,_).
T =  (A->A->A) ; T =  (A->B->A) ; T =  (A->A->B) ; T =  (A->B->B) ; 
T =  (A->B->C) ; T =  ((A->A)->A) ; T =  ((A->B)->A) ; T =  ((A->A)->B) ; 
T =  ((A->B)->B) ; T =  ((A->B)->C) .
\end{codex}
\EX
The 
sequence {\tt 2,10,75,728,8526,115764,1776060,30240210} 
counting these formulas corresponds to
the product of Catalan number of size $n$ and Bell numbers of size $n+1$,
{\tt A289679} in \cite{intseq}.

We use these formulas to test provers that show no false  negatives
on known-to-be-true formulas, as described in \cite{padl19}. The main issue
they can reveal is if they show false positives, by succeeding
on non-tautologies. This is achieved by comparing to a trusted {\em gold-standard}
 prover, e.g., one derived directly from a calculus proven sound and complete.

 \subsection{A Nested Horn Clause Tree-skeleton Generator}

The generator {\tt genHorn/3} collects leaf variables to a list, using Prolog's DCG mechanism.
\begin{code}
genHorn(N,Tree,Leaves):-genHorn(Tree,N,0,Leaves,[]).

genHorn(V,N,N)-->[V].
genHorn((A:-[B|Bs]),SN1,N3)-->{succ(N1,SN1)},[A],
  genHorn(B,N1,N2),
  genHorns(Bs,N2,N3).
  
genHorns([],N,N)-->[].
genHorns([B|Bs],SN1,N3)-->{succ(N1,SN1)},
  genHorn(B,N1,N2),
  genHorns(Bs,N2,N3).
\end{code}

\BX{generating Nested Horn Clauses}
\begin{codex}
?- genHorn(3,H,Vs).
H =  (A:-[B, C, D]), Vs = [A, B, C, D] ;
H =  (A:-[B,  (C:-[D])]), Vs = [A, B, C, D] ;
H =  (A:-[(B:-[C]), D]), Vs = [A, B, C, D] ;
H =  (A:-[(B:-[C, D])]), Vs = [A, B, C, D] ;
H =  (A:-[(B:-[(C:-[D])])]), Vs = [A, B, C, D] .
\end{codex}
\EX
Interestingly, the trees corresponding to Nested Horn Clauses are
enumerated by OEIS A000108, like purely implicational formulas,
corresponding to Catalan numbers.
Labeling of the N+1 variables serving as leaves can handled
by the same partition generator we use for labeling variables
of implicational formulas.

\subsection{Typable Closed Normal Forms of Given Size}

In direct relation to their computational uses, {\em normal forms} of simply typed
lambda terms stand out. First, this is because 
simply typed lambda terms are strongly normalizable (i.e., their
normal forms exist and are the same independently of the evaluation order).
Second, because simply-typed lambda terms share their 
most general types (called principal types) with their normal forms. Finally, 
normal forms, in combination with the right size definition \cite{maciej16},
can be described by simple CF-grammars.

Given that all formulas inhabited by a lambda terms are also inhabited by their normal forms, we can restrict ourselves to generate only typable normal forms. By generating all typable closed normal forms of a given size, we
provide known-to-be-provable formulas for the implicational fragment of
intuitionistic propositional logic.

The predicate {\tt typed\_nf/2}, given a size parameter {\tt N}, iterates, on backtracking,
over lambda terms in normal form 
{\tt X} of size {\tt N} and infers, on the fly, their type {\tt T}.
\begin{code}
typed_nf(N,X:T):-typed_nf(X,T,[],N,0).

pred(SX,X):-succ(X,SX).

typed_nf(l(X,E),(P->Q),Ps)-->pred,typed_nf(E,Q,[X:P|Ps]).  
typed_nf(X,P,Ps)-->typed_nf_no_left_lambda(X,P,Ps).

typed_nf_no_left_lambda(X,P,[Y:Q|Ps])--> agrees(X:P,[Y:Q|Ps]).
typed_nf_no_left_lambda(a(A,B),Q,Ps)-->pred,pred,
  typed_nf_no_left_lambda(A,(P->Q),Ps),
  typed_nf(B,P,Ps).

agrees(P,Ps,N,N):-member(Q,Ps),unify_with_occurs_check(P,Q).
\end{code}
As we only need the types corresponding to provable formulas,
we can omit the lambda term, resulting in a concise tautology generator
for implicational intuitionistic propositional formulas:
\begin{code}
impl_taut(N,T):-impl_taut(T,[],N,0).

impl_taut((P->Q),Ps)-->pred,impl_taut(Q,[P|Ps]).  
impl_taut(P,Ps)-->impl_taut_no_left_lambda(P,Ps).

impl_taut_no_left_lambda(P,[Q|Ps])--> agrees(P,[Q|Ps]).
impl_taut_no_left_lambda(Q,Ps)-->pred,pred,
  impl_taut_no_left_lambda((P->Q),Ps),
  impl_taut(P,Ps).  
\end{code}

\BX{Implicational tautologies, after ``numbering variables'' as natural numbers}:
\begin{code}
implTaut(N,T):-impl_taut(N,T),natvars(T). 
\end{code}

\begin{codex}
?- implTaut(4,T).
T =  (0->1->2->3->3) ;
T =  (0->1->2->3->2) ;
T =  (0->1->2->3->1) ;
T =  (0->1->2->3->0) ;
T =  (0->(0->1)->1) ;
T =  ((0->1)->0->1) ;
T =  (((0->0)->1)->1) .
\end{codex}
\EX

\BX{Counting implicational tautologies derived from typable normal forms}

\begin{codex}
?- countGen2(impl_taut,15,Rs).
Rs=[1,2,3,7,17,43,129,389,1245,4274,14991,55289,210743,826136,3354509]
\end{codex}
\EX
Note that the counts are not the same as OEIS A224345 which uses ``natural size'' of $\lambda$-terms,
 as we use here size 0 for variables, 1 for lambdas, 2 for applications.

\subsection{Generators of Canonical forms Using Commutativity, Associativity and Idempotence of Operators}

The simplest example is the  generator  {\tt allSortedHorn/2} that ``preemptively'' ensures that bodies of Nested
Horn Clauses are sorted using Prolog's standard order,
also with  duplications removed, given that conjunction is
idempotent\footnote{
\url{https://github.com/ptarau/TypesAndProofs/blob/master/allFormulas.pro}
}.
The resulting counts match {\bf A105633} in \cite{intseq}
growing with a smaller exponent than unsorted Nested Horn Clauses , which are
counted by the sequence of Catalan numbers  {\bf A000108}.

\begin{code}
genSortedHorn(N,Tree,Leaves):-succ(N,SN),length(Leaves,SN),
  generateSortedHorn(Tree,Leaves,[]).

generateSortedHorn(V)-->[V].
generateSortedHorn((A:-[B|Bs]))-->[A],
  generateSortedHorn(B),
  generateSortedHorns(B,Bs).
  
generateSortedHorns(_,[])-->[].
generateSortedHorns(B,[C|Bs])-->
  generateSortedHorn(C),
  {B@<C},
  generateSortedHorns(C,Bs).
\end{code}
This scales even more significantly in combination with a partition generator that
runs first, when more frequent identical expressions, likely to get into clause bodies are eliminated:
\begin{code}
allSortedHorn(N,T):-succ(N,SN),length(Vs,SN),
  natpartitions(Vs), % first, a partition generator
  genSortedHorn(N,T,Vs). % then, a sorted Horn clause generator
\end{code}
One can, by using the generator {\tt allStrictHorn/2},  to also eliminate the ``easy'' Horn clauses
for which the atomic head occurs in the body, to test the provers on more interesting
formulas.

Similarly, the generator {\tt genSortedTree/3}, 
in combination with a partition generator,
is used by {\tt allSortedFullFormulas/2} to
reduce equivalent formulas modulo associativity
and commutativity of conjunction and disjunction.
Other simplifications are performed at generation time, by
restricting iterated negation to at most 3, as higher number
of negations reduces to such equivalent formulas.

\subsection{Generators for ``uninhabitables''}

With help from a theorem prover, (e.g., the predicate {\tt hprove/1}) we can generate trees  that have no inhabitants for all partitions labeling their leaves as follows:

\begin{code}
unInhabitableTree(N,T):-
  genSortedHorn(N,T,Vs),
  \+ (
    natpartitions(Vs),
    hprove(T)
  ).
\end{code} 

\BX{Uninhabitable trees of size 5}
\begin{codex}
?- unInhabitableTree(5,T),nv(T).
T =  (A:-[(B:-[C]),  (D:-[E, F])]) ;
T =  (A:-[(B:-[C, D]),  (E:-[F])]) ;
T =  (A:-[(B:-[C, D, E, F])]) ;
T =  (A:-[(B:-[C, D,  (E:-[F])])]) ;
T =  (A:-[(B:-[C,  (D:-[E, F])])]) ;
T =  (A:-[(B:-[C,  (D:-[(E:-[F])])])]) ;
T =  (A:-[(B:-[(C:-[(D:-[E, F])])])]) ;
\end{codex}
\EX

We can also generate   leaf labelings such that no tree they are applied to, has inhabitants, as follows.
 
\begin{code} 
unInhabitableVars(N,Vs):-N>0,
  N1 is N-1,
  vpartitions(N,Vs),natvars(Vs),
  \+ (
    genSortedHorn(N1,T,Vs),
    hprove(T)
  ).  
\end{code}

\BX{Uninhabitable leaf labelings of size 4}
\begin{codex}
?- unInhabitableVars(4,Vs),nv(Vs).
Vs = [0, 1, 0, 0] ;
Vs = [0, 1, 1, 0] ;
Vs = [0, 1, 2, 0] ;
Vs = [0, 1, 0, 2] ;
Vs = [0, 1, 1, 1] ;
Vs = [0, 1, 2, 1] ;
Vs = [0, 1, 1, 2] ;
Vs = [0, 1, 2, 2] ;
Vs = [0, 1, 2, 3].
\end{codex}
\EX
These are dual to similar concepts investigated for lambda terms in \cite{lopstr17},
Motzkin trees that when labeled with any de Bruijn indices result in untypable terms.
Likewise, one can consider binary trees untypable with any S,K combinator labelings.

\subsection{Some Formula Count Sequences for Small Sizes}

By counting the number of solutions of our generators by increasing sizes, we obtain
some interesting formula counts. We list them here together with the names of the predicates
that given N as their first argument return the list of count up to N as their second argument.

\BI
\I countHornTrees = A000108: Catalan numbers 1, 2, 5, 14, 42, 132, 429, 1430, 4862


\I countSortedHorn = A105633:  1, 2, 4, 9, 22, 57, 154, 429, 1223, 3550, 10455, 31160, 93802, 284789

\I countHorn3 = NEW: 1, 1, 2, 5, 13, 37, 109, 331, 1027, 3241, 10367, 33531, 109463

\I countSortedHorn3=NEW: 1, 2, 4, 8, 20, 47, 122, 316, 845, 2284, 6264, 17337, 48424, 136196, 385548

\I all implicational intuitionistic propositional calculus formulas = A289679: 1,  2,  10,  75,  728,  8526,  115764,  1776060,  30240210
\I all provable implicational intuitionistic propositional calculus formulas = NEW: 0, 1, 3, 24, 201, 2201, 27406, 391379, 6215192

\I countUnInhabitableTree = NEW: 1, 0, 1, 1, 4, 7, 23, 53, 163, 432, 1306

\I countUnInhabitableVars = NEW: 0, 1, 1, 4, 9, 30, 122, 528, 2517, 12951, 71455
\EI

\section{Random Formula Generation Algorithms}\label{rand}

An advantage of random formulas of size much larger than those generated by an exhaustive enumeration at a given size, is that such formulas can be potentially harder for the provers, reveal phenomena not present at smaller sizes (e.g., unexpected space complexity), and more generally, they can test for scalability issues.

\subsection{Random Simply-typed Terms, with Boltzmann Samplers}

Once passing correctness tests,  our provers need to be tested against
large random terms. The mechanism is similar to the use of all-term
generators.

We generate random simply-typed normal forms, using a Boltzmann sampler
along the lines of that described in \cite{tplp18}. The code variant, adapted to
our different term-size definition is at:\\
{\small \url{https://github.com/ptarau/TypesAndProofs/blob/master/ranNormalForms.pro}}.
It works as follows:
\begin{codex}
?- ranTNF(60,XT,TypeSize).
XT = l(l(a(a(0, l(a(a(0, a(0, l(...))), s(s(0))))), 
             l(l(a(a(0, a(l(...), a(..., ...))), l(0))))))) 
        : 
        (A->((((A->A)- ...)->D)->D)->M)->M),
TypeSize = 34.
\end{codex}
Interestingly, partly due to the fact that there's some variation in the size of the terms
that Boltzmann samplers generate, and more to the fact that the distribution of 
types  favors (as seen in the second example) the simple tautologies
where an atom identical to the last one is contained in the implication 
chain leading to it \cite{tautintclass,density08}, 
if we want to use these for scalability tests, additional filtering
mechanisms need to be used to statically reject  type
expressions that are large but easy to prove as intuitionistic tautologies.

\subsection{Random Implicational Formulas}
The generation of random implicational formulas relies on a random binary tree generator, combined with
a random set partition generator.

Our code  combines an implementation of R\'emy's algorithm \cite{remy85}, 
along the lines
of Knuth's algorithm {\bf R} in \cite{knuth_trees}  for the {\em generation of
random binary} trees at\\
{\small \url{https://github.com/ptarau/TypesAndProofs/blob/master/RemyR.pro}}
with code to generate {\em random set partitions} at:\\
{\small \url{https://github.com/ptarau/TypesAndProofs/blob/master/ranPartition.pro}}.

We refer to \cite{sac18} for a declarative implementation of R\'emy's algorithm in Prolog with code
adapted for this paper at:\\
{\small \url{https://github.com/ptarau/TypesAndProofs/blob/master/RemyP.pro}}.

As automatic Boltzmann sampler generation of set partitions 
is limited to fixed numbers
of equivalence classes from which a CF- grammar can be given,
we build our  random set partition generator that  groups 
variables in leaf position into equivalence classes by using an 
urn-algorithm \cite{stam_urn}. 
Once a random binary tree of size $N$ is generated with the \verb~->/2~ constructor labeling internal nodes,
the $N+1$ leaves of the tree are decorated with variables denoted by successive integers
starting from 0. As variables sharing a  name
define equivalence classes on the set of variables, each choice of them
 corresponds to a set partition of the $N+1$ nodes.
 Thus, a set partition of the leaves
{\tt \{0,1,2,3\}} like {\tt
\{\{0\},\{1,2\},\{3\}\}} will  correspond to the variable
leaf decorations {\tt \[0,1,1,2\]}
The partition generator works as follows:
\begin{codex}
?- ranSetPart(7,Vars).
Vars = [0, 1, 2, 1, 1, 2, 3] .
\end{codex}
Note that the list of labels it generates can be directly used to decorate
the random binary tree generated by R\'emy's algorithm, by unifying
the list of variables {\tt Vs} with it.
\begin{codex}
?- remy(6,T,Vs).
T =  ((((A->B)->C->D)->E->F)->G),
Vs = [A, B, C, D, E, F, G] .
\end{codex}

The combined generator, that produces in a few seconds terms of size 1000,
works as follows:

\begin{codex}
?- time(ranImpFormula(1000,_)). 
% includes tabling large Stirling numbers
% 37,245,709 inferences,7.501 CPU in 
7.975 seconds (94% CPU, 4965628 Lips)

?- time(ranImpFormula(1000,_)). % fast, thanks to tabling
% 107,163 inferences,0.040 CPU in 
0.044 seconds (92% CPU, 2659329 Lips)
\end{codex}.
Note that we use Prolog's {\em tabling} (a form of automated dynamic programming)
to avoid costly recomputation of the (very large) Sterling numbers in the code at:\\
{\small \url{https://github.com/ptarau/TypesAndProofs/blob/master/ranPartition.pro}}.

\subsection{Generating Random Tautologies from Typable Combinator Expressions}

Boltzmann samplers for the uniform random generation of 
 simply typed lambda terms and their normal forms
are described in \cite{tplp18}. Of particular interest for their use as
generators of random intuitionistic tautologies are the types of the terms in normal form,
as every type inferred for a simply typed term can also be obtained after
$\beta$-reduction, from its normal form. Thus, we can work on a smaller set of terms
while obtaining the same set of formulas\footnote{See \url{https://github.com/ptarau/TypesAndProofs/blob/master/ranNormalForms.pro}}.

One might ask why not use  Hilbert-style axioms with substitutions and modus ponens
to generate directly provable formulas. After all, these axioms are simple enough 
and exactly mimic the types of the {\tt S} and {\tt K} combinators:\\

\noindent
\begin{math}
K: ~~~~~ A\rightarrow\left(B\rightarrow A\right)\\
S: ~~~~~ \left(A\rightarrow\left(B\rightarrow C\right)\right)\rightarrow \left(\left(A\rightarrow B\right)\rightarrow\left(A\rightarrow C\right)\right)\\
\end{math}

From them, we derive new theorems 
by applying substitution of formulas for variables in axioms and theorems and by
applying the modus ponens inference rule:\\\\
$MP: ~~~~~ A,~A \rightarrow B ~\vdash~ B$.\\

In fact, doing so would bring us back in time to the 1930's,  before Gentzen's work on using sequent calculus for deduction \cite{gent} !
The problem is that while applying substitutions to the axioms and theorems
is fairly simple (especially with Prolog's logic 
variables), finding the two already
known theorems needed to activate modus ponens in a growing stream of theorems
is computationally prohibitive.
A better (implicit) use of the $S$ and $K$ axioms is by designing a generator for simply typed {\tt SK}-expressions. With them, one  obtains the same set of the types/tautologies as those generated for simply typed lambda terms using Boltzmann samplers, given that all lambda terms are expressible as combinator formulas.

Our implementation\footnote{at \url{https://github.com/ptarau/TypesAndProofs/blob/master/RemyR.pro}}
 generates uniformly random binary trees of a given size using  R\'emy's algorithm \cite{remy85}, along the lines
of Knuth's algorithm {\bf R} in \cite{knuth_trees}, with leaves decorated with randomly selected symbols from the set {\tt \{s,k\}}. A declarative Prolog implementation of R\'emy's algorithm is described  in \cite{sac18}. We have also adapted its code\footnote{at
\url{https://github.com/ptarau/TypesAndProofs/blob/master/RemyP.pro}},
as a somewhat slower alternative to Knuth's algorithm. Using Knuth's algorithm {\bf R}, the predicate {\tt remy\_sk/2}  generates in a few seconds random {\tt SK}-trees with 2-3 million nodes.

The predicate {\tt ranSK/3} filters the random {\tt SK}-trees of size {\tt N}
to represent typable combinator expressions,
while ensuring that their types are of size at least {\tt M}, to avoid the frequently
occurring trivial types.
\begin{code}
ranSK(N,M,T):-
  repeat,
    % generates a binary expression tree with S or K at its leaves
    remy_sk(N,X),
    sk_type_of(X,T), % infers the type of an SK-expression
    tsize(T,S), % computes the size of the inferred type
    S>=M,
  !,
  natvars(T). % binds type variables to natural numbers, starting from 0
\end{code}
The type inference algorithm for {\tt SK}-expressions is quite simple.
After stating that {\tt s} and {\tt k} leaves are well typed, we ensure
that the types of application nodes agree (as in the modus-ponens rule),
using sound unification to avoid creation of cyclical type formulas.
\begin{code}
sType((A->B->C)->(A->B)->A->C).

kType((A->_B->A)).
  
sk_type_of(s,T):-sType(T). % S-leaf's type
sk_type_of(k,T):-kType(T). % K-leaf's type
sk_type_of((A*B),Target):- % application node
  sk_type_of(A,SourceToTarget),
  sk_type_of(B,Source),
  unify_with_occurs_check(SourceToTarget,(Source->Target)). 
\end{code}

\BX
Large random implicational tautologies, comparable to those generated using
 Boltzmann samplers, can be produced in a few seconds
by inferring the types
of random {\tt SK}-expressions.
\begin{codex}
?- ranSK(60,40,T).
T =  (((((0->((1->2->3)->(1->2)->1->3)->4)->0)->0->((1->2->3)->
 (1->2)->1->3)->4)->(0->((1->2->3)->(1->2)->1->3)->4)->0)->
 (((0->((1->2->3)->(1->2)->1->3)->4)->0)->0->((1->2->3)->(1->2)->1->3)->4)->
 ((1->2->3)->(1->2)->1->3)->4).
\end{codex}
\EX


\section{Formula Transformers Reducing Test Sets to Canonical Representatives of Equivalence Classes}\label{trans}

Formula transformers serve several purposes. First, we want to perform simplifications to facilitate the work of the provers. This is achieved by converting between equivalent representations w.r.t. provability.
Second, the tautologies we generate might be too easy for the provers and we can, by applying the transformations in reverse,  make them significantly harder, while still knowing their status as being provable or not.

Conversely, relying on provers known to be sound and complete,
we can establish correctness of the transformers as agreement on the success of a correct prover before and after a transformation is applied.

\subsection{The Mints Transformation}

Grigori Mints has proven, in his seminal paper studying complexity classes for 
intuitionistic propositional logic \cite{mints92}, that a formula $f$
is equiprovable to a formula of the form $X_f \rightarrow g$ where
$X_f$ is a conjunction of formulas of one of the forms
$p, \char`\~p,~ p \rightarrow q,~(p \rightarrow q)  \rightarrow r,~ p \rightarrow (q \rightarrow r),~  p \rightarrow (q ~v~ r),~ p \rightarrow  \char`\~q,~  \char`\~q \rightarrow p $. With introduction of new variables (like with the Tseitin transform for SAT or ASP solvers), the transformation is linear in space.

We have implemented a variant of the Mints transformation\footnote{
\url{https://github.com/ptarau/TypesAndProofs/blob/master/mints.pro}
}
that also eliminates negation by replacing \verb|~p| with \verb~p->false~ and expands
the equivalence relation ``\verb~<->~''.
The correctness of our implementation has been tested by showing that
on formulas of small sizes, a trusted prover succeeds on the same set of formulas
before and after the transformation.
As transforming formulas known-to-be-true results in formulas of a larger size, 
we have  used them as scalability tests for the provers.
For disjunction-free formulas, in combination with a converter to Nested Horn Clause form,
the transformation has been used to generate equivalent 
Nested Horn Clauses of depth at most 3,
a new canonical form, also useful for scalability tests for our provers.

\subsection{Transforming Disjunction-free Propositional Formulas to Lists of Nested Horn Clauses}

The predicate {\tt toNestedHorn/2} transforms a disjunction-free 
propositional formula to a Nested Horn Clause form, which is essentially
the same as the language of propositional N-Prolog \cite{nprolog84}.
\begin{code}
toNestedHorn(A,R):-
  expand_equiv(A,X),toHorn1(X,H),expand_horn(H,E),reduce_heads(E,R). 
\end{code}
After expanding equivalences \verb~A<->B~ to conjunctions of implications it
converts chained implications to Horn clauses, flattens conjunctions in their
bodies and reduces formulas in head positions until  all heads are atomic\footnote{
\url{https://github.com/ptarau/TypesAndProofs/blob/master/toHorn.pro}
}.
The resulting formulas can then be proven or refuted by invoking 
a Nested Horn Clause prover (like {\tt ahprove/1}, to be described in section \ref{use})
on each member of 
the list of nested clauses. This will result in reducing worst case
 space complexity from exponential to
 $O(n~log(n))$.
\BX
Expansion to equivalent set of Nested Horn Clauses.
\begin{codex}
?-toNestedHorn(a&b&(c&d->e)<->f&g,R).
R = [(f:-[a,b,(e:-[c,d])]),(g:-[a,b,(e:-[c,d])]),(a:-[f,g]),
     (b:-[f,g]),(e:-[c,d,f,g])].
\end{codex}
\EX

\subsection{Transforming to the disjunction-biconditional-negation base}

An alternative base for intuitionistic propositional logic is the one consisting
of disjunction, biconditional and negation. Implication and conjunction can be expressed in terms of them as follows.
\begin{code}
toDisjBiCond((A->B),R):-!,toDisjBiCond(A,X),toDisjBiCond(B,Y),
          R=((X v Y)<->Y).
toDisjBiCond(A & B,R):-!,toDisjBiCond(A,X),toDisjBiCond(B,Y),
          R=((X v Y)<->(X<->Y)).
toDisjBiCond(A v B,R):-!,toDisjBiCond(A,X),toDisjBiCond(B,Y),
          R=(X v Y).
toDisjBiCond(A<->B,R):-!,toDisjBiCond(A,X),toDisjBiCond(B,Y),
          R=(X<->Y).
toDisjBiCond(~A,R):-!,toDisjBiCond(A,X),
         R = (~X).
toDisjBiCond(A,A).
\end{code}
This makes formulas larger and much harder to solve, especially as biconditional ``\verb~<->~'' is expanded to  a conjunction of implications. Note that the reverse of the transformation
actually works as a good simplifier for formulas passed to the provers.

\section{Deriving Lightweight Theorem Provers for Intuitionistic Propositional Logic}\label{use}

Initially, like for other fields of mathematics and logic,
Hilbert-style axioms were considered for intuitionistic logic. While simple and directly mapped
to SKI-combinators via the Curry-Howard isomorphism, their usability for automation is very
limited. In fact, their inadequacy for formalizing even "hand-written" mathematics
was the main trigger of Gentzen's work on natural deduction and sequent calculus,
inspired by the need for formal reasoning in the foundation of mathematics \cite{gent}.

Thus, we start with Gentzen's own calculus for intuitionistic logic, simplified here
to only cover the purely implicational fragment, given that our focus is on theorem
provers working on formulas that  correspond to types of  simply-typed lambda terms.

\subsection{Gentzen's LJ Calculus, Restricted to the Implicational Fragment of Propositional Intuitionistic Logic}

We assume familiarity with basic sequent calculus notation.
Gentzen's  original LJ calculus \cite{gent} (with the equivalent notation of \cite{dy1}) uses the following rules.\\\\

{\Large
\noindent
\begin{math}
LJ_1:~~~~\frac{~}{A,\Gamma ~\vdash~ A}\\\\\\
LJ_2:~~~~\frac{A,\Gamma ~\vdash~ B}{\Gamma ~\vdash~ A\rightarrow B}\\\\\\
LJ_3:~~~~\frac{A \rightarrow B,\Gamma ~\vdash~ A ~~~~ B,\Gamma ~\vdash~ G}
{A \rightarrow B, \Gamma ~\vdash~ G}\\
\end{math}
}

As one can easily see, when trying a goal-driven implementation that uses the rules in upward direction, the unchanged premises on left side of rule $LJ_3$ would not ensure termination as nothing prevents $A$ and $G$ from repeatedly trading places during the inference process.


A good starting point for developing heuristic-free, lightweight provers is to directly derive them from calculi that have been proven sound and complete.

\subsection{The LJT/G4ip Calculus, Restricted to the Implicational Fragment}

Motivated by problems related to loop avoidance in implementing  Gentzen's {\bf LJ} calculus,
Roy Dyckhoff \cite{dy1}  introduces
the following rules for his {\tt LJT} calculus\footnote{Also called the G4ip calculus. Restricted here to the implicational fragment.}.\\\\
{\large
\noindent
\begin{math}
LJT_1:~~~~\frac{~}{A,\Gamma ~\vdash~ A}\\\\\\
LJT_2:~~~~\frac{A,\Gamma ~\vdash~ B}{\Gamma ~\vdash~ A\rightarrow B}\\\\\\
LJT_3:~~~~\frac{B,A,\Gamma ~\vdash~ G}{A \rightarrow B,A,\Gamma ~\vdash~ G}\\\\\\ 
LJT_4:~~~~\frac{D \rightarrow B,\Gamma ~\vdash~ C \rightarrow D ~~~~ B,\Gamma ~\vdash~ G}
{ \left( C \rightarrow D \right) \rightarrow B,\Gamma ~\vdash~ G }\\
\end{math}
}\\
The rules work with the context $\Gamma$
being either a multiset or a set.\\\\

In \cite{padl19}, the following literal translation of the rules $LJT_1 \ldots LJT_4$ to Prolog is given, with values in the environment $\Gamma$ denoted by the variable {\tt Vs}.

\begin{code}
lprove(T):-ljt(T,[]).

ljt(A,Vs):-memberchk(A,Vs),!.       % LJT_1
ljt((A->B),Vs):-!,ljt(B,[A|Vs]).    % LJT_2 
ljt(G,Vs1):- %atomic(G),            % LJT_3
  select((A->B),Vs1,Vs2),
  memberchk(A,Vs2),!,
  ljt(G,[B|Vs2]).
ljt(G,Vs1):-                        % LJT_4
  select( ((C->D)->B),Vs1,Vs2),
  ljt((C->D), [(D->B)|Vs2]),!,
  ljt(G,[B|Vs2]).
\end{code}
Note the use of {\tt select/3} to extract a term from the environment (a nondeterministic step).
The advantage of these rules is that they do not need loop checking to ensure termination, 
as one can identify a multiset ordering-based size definition that decreases after each step \cite{dy1}.

Next, we will show   provers derived from {\tt lprove/1} via refinements
validated by our testing framework, among which ones 
that reduce the  exponential worst case space complexity of 
{\tt lprove/1} to $O(n~log(n))$.

\subsection{Concentrating Nondeterminism into One Place}

We start with a transformation that keeps the underlying implicational formula unchanged.
It  merges the work of the two {\tt select/3} calls
into a single call, observing that their respective clauses
 do similar things after the call to {\tt select/3}.
That avoids redoing the same iteration over candidates for reduction.

\begin{code}
bprove(T):-ljb(T,[]),!.

ljb(A,Vs):-memberchk(A,Vs),!.
ljb((A->B),Vs):-!,ljb(B,[A|Vs]). 
ljb(G,Vs1):-
  select((A->B),Vs1,Vs2),
  ljb_imp(A,B,Vs2),
  !,
  ljb(G,[B|Vs2]).

ljb_imp((C->D),B,Vs):-!,ljb((C->D),[(D->B)|Vs]).
ljb_imp(A,_,Vs):-atomic(A),memberchk(A,Vs). 
\end{code}

\subsection{Implicational Formulas as Nested Horn Clauses}

Given the equivalence between:
$B_1 \rightarrow B_2 \ldots B_n \rightarrow H$ and (in Prolog notation)
$ H$ {\tt :-} $B_1,B_2, \ldots, B_n$,
(where we choose $H$ as the {\em atomic} formula ending a chain
of implications), we can, recursively, transform an implicational formula
into one built form nested clauses, as follows.

\begin{code}
toHorn((A->B),(H:-Bs)):-!,toHorns((A->B),Bs,H).
toHorn(H,H).

toHorns((A->B),[HA|Bs],H):-!,toHorn(A,HA),toHorns(B,Bs,H).
toHorns(H,[],H).    
\end{code}
Note also that the transformation is reversible and that lists 
(instead of Prolog's conjunction chains)
are used to collect the elements of the body of a clause.

\begin{codex}
?- toHorn(((0->1->2->3->4)->(0->1->2)->0->2->3),R).
R =  (3:-[(4:-[0, 1, 2, 3]),  (2:-[0, 1]), 0, 2]).
\end{codex}

This suggests transforming provers for implicational formulas into
equivalent provers working on nested Horn clauses.
\begin{code}
hprove(T0):-toHorn(T0,T),ljh(T,[]),!.

ljh(A,Vs):-memberchk(A,Vs),!. 
ljh((B:-As),Vs1):-!,append(As,Vs1,Vs2),ljh(B,Vs2).
ljh(G,Vs1):-               % atomic(G), G not in Vs1
  memberchk((G:-_),Vs1),   % if not, we just fail!
  select((B:-As),Vs1,Vs2), % outer select loop
  select(A,As,Bs),         % inner select loop
  ljh_imp(A,B,Vs2),        % A is an element of the body of B
  !,
  trimmed((B:-Bs),NewB),   % trim off empty bodies
  ljh(G,[NewB|Vs2]).
  
ljh_imp((D:-Cs),B,Vs):-!,ljh((D:-Cs),[(B:-[D])|Vs]).
ljh_imp(A,_B,Vs):-memberchk(A,Vs).

trimmed((B:-[]),R):-!,R=B.
trimmed(BBs,BBs).
\end{code}
A first improvement, ensuring quicker rejection of non-theorems is
the call to {\tt memberchk} in the 3-rd clause to ensure
that our goal {\tt G} is the head of at least one of the assumptions.
Once that test is passed, the 3-rd clause works as a reducer
of the assumed hypotheses. It removes from the context a clause \verb~B:-As~
and it removes from its body a formula \verb~A~, to be passed
to \verb~ljh_imp~, with the remaining context.
Should \verb~A~ be atomic, we succeed if and only
if  it is already in the context.
Otherwise, we closely mimic rule $LJT_4$ by trying
to prove \verb~A = (D:-Cs)~, after extending the context
with the assumption \verb~B:-[D]~.
Note that in both cases the context gets smaller,
as \verb~As~ does not contain the \verb~A~ anymore.
Moreover, should the body \verb~Bs~ end up empty, the
clause is downgraded to its atomic head by the predicate {\tt trimmed/2}.
Also, by having a second {\tt select/3} call
in the third clause of {\tt ljh}, will give {\tt ljh\_imp} more
chances to succeed and commit.

Thus, besides quickly filtering out failing search branches,
the nested Horn clause form of implicational
logic helps bypass some intermediate steps, by focusing
on the head of the Horn clause, which corresponds to the
last atom in a chain of implications.

\subsection{Propagating Back the Elimination of Non-matching Heads} 

We can propagate back to the implicational forms used in {\tt bprover} 
the observation made on the Horn-clause form that
heads (as computed below) should match at least one assumption.
\begin{code}
head_of(_->B,G):-!,head_of(B,G).
head_of(G,G). 
\end{code}

We can apply this to {\tt bprove/1}
as shown in the 3-rd clause of {\tt lje}, where we can also
prioritize the assumption found to have the head {\tt G}, 
by placing it first in the context.

\begin{code}
eprove(T):-lje(T,[]),!.

lje(A,Vs):-memberchk(A,Vs),!.
lje((A->B),Vs):-!,lje(B,[A|Vs]). 
lje(G,Vs0):-
  select(T,Vs0,Vs1),head_of(T,G),!,
  select((A->B),[T|Vs1],Vs2),lje_imp(A,B,Vs2),!,
  lje(G,[B|Vs2]).

lje_imp((C->D),B,Vs):-!,lje((C->D),[(D->B)|Vs]).
lje_imp(A,_,Vs):-atomic(A),memberchk(A,Vs).
\end{code}

This brings the performance of {\tt eprove} within a few percents of {\tt hprove}. 
 
\subsection{Extracting the Proof Terms}

Extracting the {\em proof terms} (lambda terms having the formulas we prove as types) is achieved
by decorating in the code  with application nodes {\tt a/2},
lambda nodes {\tt l/2} (with first argument a logic variable)
and leaf nodes (with logic variables, same as the identically 
named ones  in the first argument of the corresponding {\tt l/2} nodes).

The simplicity of the predicate {\tt eprove/1} and the fact that
this is essentially the inverse of a type inference  algorithm (e.g., the one in \cite{hiking17})
point out how the decoration mechanism works.

\begin{code}
sprove(T):-ljs(X,T,[]).

ljs(X,A,Vs):-memberchk(X:A,Vs),!. % leaf variable
ljs(l(X,E),(A->B),Vs):-!,ljs(E,B,[X:A|Vs]).  % lambda term
ljs(E,G,Vs1):- 
  member(_:V,Vs1),head_of(V,G),!, % fail if non-tautology
  select(S:(A->B),Vs1,Vs2),       % source of application
  ljs_imp(T,A,B,Vs2),             % target of application
  !,
  ljs(E,G,[a(S,T):B|Vs2]).        % application
  
ljs_imp(l(X,E),(C->D),B,Vs):-!,ljs(E,(C->D),[X:(D->B)|Vs]).
ljs_imp(E,A,_,Vs):-memberchk(E:A,Vs). 
\end{code}

Thus, lambda nodes decorate  {\em implication introductions} and
application nodes  decorate {\em modus ponens} reductions
in the corresponding calculus. Note that the two clauses of
{\tt  ljs\_imp} provide the target node $T$. When seen from
the type inference side, $T$ is the type resulting
from cancelling the source type $S$ and
the application type $S \rightarrow T$.

Calling {\tt sprove/2} on the formulas corresponding to the types
of the $S, K$ and $I$ combinators, we obtain:
\begin{codex}
?- sprove(((0->1->2)->(0->1)->0->2),X). 
X = l(A, l(B, l(C, a(a(A, C), a(B, C))))).                % S
?- sprove((0->1->0),X).
X = l(A, l(B, A)).                                        % K
?- sprove((0->0),X).
X = l(A, A).                                              % I
\end{codex}

\subsection{A $O(n~log(n))$ Space Complexity Prover Implementing Hudelmaier's Calculus}

In \cite{hud93} a sequent calculus for intuitionistic propositional logic 
ensuring $O(n~log(n))$ space complexity is introduced. We have implemented its restriction to the implicational subset as the predicate {\tt nvprove/1}, derived in a few simple steps from {\tt lprove/1}. The new variables are introduced by
using a DCG transformation that advances a variable counter starting at {\tt 10000}.
\begin{code}
nvprove(T):-ljnv(T,[],10000,_).

ljnv(A,Vs)-->{memberchk(A,Vs)},!.
ljnv((A->B),Vs)-->!,ljnv(B,[A|Vs]). 
ljnv(G,Vs1)--> % atomic(G),
  {select((A->B),Vs1,Vs2)},
  ljnv_imp(A,B,Vs2),
  !,
  ljnv(G,[B|Vs2]).

ljnv_imp((C->D),B,Vs)-->!,newvar(P),ljnv(P,[C,(D->P),(P->B)|Vs]).
ljnv_imp(A,_,Vs)-->{memberchk(A,Vs)}.  

newvar(N,N,SN):-succ(N,SN).
\end{code}

Hudelmaier's algorithm achieves $O(n~log(n))$ worst case space complexity by avoiding to duplicate the possibly large subterm
{\tt D} in rule $LJT_4$. After proving that this transformation results in a tautology if and only if the original term was provable, instead of  duplicating  {\tt D} in the last clause of {\tt ljt/2},  Hudelmaier introduces a new variable {\tt P}, that we implement using the DCG step {\tt newvar/3} 
in the first clause of {\tt ljnv\_imp/5}.

\subsection{A $O(n*log(n))$ Space Complexity  Nested Horn Clause Prover }

After the transformation steps shown for {\tt hprove}, that use the fact
that $a_1 \rightarrow a_2 \ldots  \rightarrow a_n \rightarrow  a_0$ is equivalent to
  $a_0 \leftarrow a_1 ~\&~ a_2 ~\&~ \ldots ~\&~ a_n$ and  elimination of  the interpreter wrapper by defining the predicate ``{\tt <-}'' directly, we activate the proof with {\tt call(H)}, after using the transformer {\tt toAHorn/2} to convert our tests from their implicational form to an equivalent Nested Horn Clause form.
\begin{code}
ahprove(A):-toAHorn(A,H),call(H).
\end{code}

Then, the algorithm proceeds by reducing the uniformly represented Nested Horn Clauses
 of the form {\tt Head <- ListOfBodyTerms}. Note also that  the sequent-calculus form
 is not used anymore as a meta-rule, as it can be, equivalently, folded into a Nested Horn Clause form.
\begin{code}
:-op(800,xfx,(<-)).

A<-Vs:-memberchk(A,Vs),!. 
(B<-As)<-Vs1:-!,append(As,Vs1,Vs2),B<-Vs2.

G<-Vs1:- % atomic(G), G not on Vs1
  memberchk((G<-_),Vs1), % if not, we just fail
  select(B<-As,Vs1,Vs2), % outer select loop
  select(A,As,Bs),         % inner select loop
  ahlj_imp(A,B,Vs2), % A element of the body of B
  !,
  atrimmed(B<-Bs,NewB), % trim empty bodies
  G<-[NewB|Vs2].
  
ahlj_imp(D<-Cs,B,Vs):-!, (D<-Cs)<-[B<-[D]|Vs].
ahlj_imp(A,_B,Vs):- memberchk(A,Vs).

atrimmed(B<-[],R):-!,R=B. 
atrimmed(BBs,BBs).
\end{code}

A few words on the {\em story} that got us here. We have  observed 
that the  Nested Horn Clause prover {\tt hprove/1}  outperforms other provers (e.g., {\tt bprove/1} by more than an order of magnitude (e.g., {\bf 121.006} seconds vs. {\bf 3221.227} seconds on terms of size {\bf 16}).

But, we have not had  a convincing explanation why this is the case.
The fact that a test-driven refinement step implementing Hudelmaier's introduction of auxiliary variables brought our implication-based prover  much closer in performance to the Nested Horn Clause transform, hinted towards the fact that  that Hudelmaier's optimization shares a relevant similarity with the Nested Horn Clause prover.
Finally, it  became clear that the duplicated formula {\tt D} in {\tt ahlj\_imp/3}, as it occurs as the head of a clause, is atomic in the Nested Horn Clause prover and thus the space  increase is bounded by the number of atoms in the original formula to be proven, without the need for introducing new variables.

\subsection{A Lightweight Theorem Prover for Full Intuitionistic Propositional Logic}

Starting from the sequent calculus for the full intuitionistic propositional logic in LJT/G4ip \cite{dy1}, to
which we have also added rules for the ``\verb~<->~'' relation, we obtain the following lightweight prover.
\begin{codeh}
:- op(425,  fy,  ~ ).
:- op(450, xfy,  & ).    
:- op(475, xfy,  v ).    
:- op(500, xfx,  <-> ).  
\end{codeh}

\begin{code}
ljfa(T):-  ljfa(T,[]).

ljfa(A,Vs):-memberchk(A,Vs),!.
ljfa(_,Vs):-memberchk(false,Vs),!.
ljfa(A<->B,Vs):-!,ljfa(B,[A|Vs]),ljfa(A,[B|Vs]).
ljfa((A->B),Vs):-!,ljfa(B,[A|Vs]).
ljfa(A & B,Vs):-!,ljfa(A,Vs),ljfa(B,Vs).
ljfa(G,Vs1):- % atomic or disj or false
  select(Red,Vs1,Vs2),
  ljfa_reduce(Red,G,Vs2,Vs3),
  !,
  ljfa(G,Vs3).
ljfa(A v B, Vs):-(ljfa(A,Vs);ljfa(B,Vs)),!.

ljfa_reduce((A->B),_,Vs1,Vs2):-!,ljfa_imp(A,B,Vs1,Vs2).
ljfa_reduce((A & B),_,Vs,[A,B|Vs]):-!.
ljfa_reduce((A<->B),_,Vs,[(A->B),(B->A)|Vs]):-!.
ljfa_reduce((A v B),G,Vs,[B|Vs]):-ljfa(G,[A|Vs]).
  
ljfa_imp((C->D),B,Vs,[B|Vs]):-!,ljfa((C->D),[(D->B)|Vs]).
ljfa_imp((C & D),B,Vs,[(C->(D->B))|Vs]):-!.
ljfa_imp((C v D),B,Vs,[(C->B),(D->B)|Vs]):-!.
ljfa_imp((C<->D),B,Vs,[((C->D)->((D->C)->B))|Vs]):-!.
ljfa_imp(A,B,Vs,[B|Vs]):-memberchk(A,Vs).  
\end{code}
 We
validate it first by testing it on the implicational subset, then against Roy Dyckhoff's Prolog implementation\footnote{
\url{https://github.com/ptarau/TypesAndProofs/blob/master/third_party/dyckhoff_orig.pro}
},
working on formulas generated by the predicate {\tt allSortedFullFormulas/2} up to size 12. Finally we run it
on the human-made tests at \url{http://iltp.de} on which we get no errors, solving correctly 161 problems, with a 60 seconds timeout, compared with the
175 problems solved by Roy Dyckhoff's heuristics-based 400 lines prover,
with the same timeout\footnote{
\url{https://github.com/ptarau/TypesAndProofs/blob/master/tester.pro}
}.
On the other hand, the performance of {\tt ahprove/1} is significantly better
when compared with the iLeanTap \cite{otten97}, a 122 lines ``lean'' theorem prover\footnote{
\url{https://github.com/ptarau/TypesAndProofs/blob/master/third_party/ileantap.pro}
}  that only solves 35 problems correctly and makes 3 errors with the same
60 seconds timeout.

Among its applications, is a derivation of an embedding of Artemov and Protopopescu's Intuitionistic Epistemic Logic
\cite{iel16} in Intuitionistic Propositional Logic \cite{ieldefs}, where this prover is used as an oracle for candidate definitions for epistemic operators for which theorems of the logic should hold and non-theorems fail.

\section{The Testing Framework}\label{tests}

Correctness can be checked by identifying false positives or false negatives. A false positive is  a non-tautology that the prover proves, breaking the {\em soundness} property.
A false negative is a tautology that the prover fails to prove,
breaking the {\em completeness} property.
While classical tautologies are easily tested (at small scale against truth tables, at medium scale with 
classical propositional provers and at larger scale with a SAT solver), intuitionistic provers require a more creative approach, given the absence of a finite truth-value table model.

As a first bootstrapping step, assuming that no "gold standard" prover is available, one can look at the other side of the Curry-Howard isomorphism, and rely on generators of (typable) lambda terms and generators implicational logic formulas, with results being checked against a trusted type inference algorithm.

As a next step, a trusted prover can be used as a ``gold standard'' to test both for false positives and negatives.

\subsection{Finding False Negatives by Generating the Set of Simply Typed Normal Forms of a Given Size}

A false negative is identified if our prover  fails on a type expression known to have an inhabitant.
Via the Curry-Howard isomorphism, such terms are the types inferred for lambda terms,
generated by increasing sizes. In fact, this means that all implicational formulas
having proofs shorter than a given number are all covered, but possibly
small formulas having long proofs might not be reachable with this method
that explores the search by the size of the proof rather than the size
of the formula to be proven.
We refer to \cite{hiking17} for a detailed description of efficient algorithms 
generating pairs of simply typed lambda terms in normal form together with their principal types.
The  code we use here is at: 
{\small  \url{https://github.com/ptarau/TypesAndProofs/blob/master/allTypedNFs.pro}}

\subsection{Finding False Positives by Generating All 
Implicational Formulas/Type Expressions of a Given Size}

A false positive is identified if the prover succeeds finding an inhabitant
for a type expression that does not have one.
 
We obtain type expressions by generating all binary trees of a given size, extracting their
leaf variables and then iterating over the set of their set partitions, while
unifying variables belonging to the same partition. 
We refer to \cite{hiking17} for a detailed description of the algorithms.

The code describing the all-tree and set partition generation as well as their integration as a type expression generator is at:\\
{\small \url{https://github.com/ptarau/TypesAndProofs/blob/master/allPartitions.pro}}.

We have tested the predicate {\tt lprove/1} as well as all other provers derived from it
for false negatives 
against simple types of terms up to size 15 (with size defined as 2 for
applications, 1 for lambdas and 0 for variables) and for false positives against
all type expressions up to size 7 (with size defined as the number of internal nodes).

An advantage of exhaustive testing with all formulas of a given size is that it implicitly ensures coverage: no path is missed simply because there are no paths left unexplored.

\subsection{Testing Against a Trusted Reference Implementation}

Assuming we trust an existing reference implementation (e.g., after it passes our generator-based tests),
it makes sense to use it
as a "gold standard". In this case, we can identify both false positives and negatives
directly, as follows:

\begin{code}
gold_test(N,Generator,Gold,Silver, Term, Res):-call(Generator,N,Term),
  gold_test_one(Gold,Silver,Term, Res),
  Res\=agreement.
  
gold_test_one(Gold,Silver,T, Res):-
  ( call(Silver,T) -> \+ call(Gold,T), 
    Res = wrong_success
  ; call(Gold,T) -> % \+ Silver
    Res = wrong_failure
  ; Res = agreement
  ).
\end{code}

When specializing to a generator for all well-formed implication expressions,
and using Dyckhoff's dprove/1 predicate as a gold standard, we have:

\begin{code}
gold_test(N, Silver, Culprit, Unexp):-
  gold_test(N,allImpFormulas,dprove,Silver,Culprit,Unexp).
\end{code}

To test the tester, we design a prover that randomly succeeds or fails.

\begin{code}
badProve(_) :- 0 =:= random(2).
\end{code}

We can now test lprove/1 and badprove/1 as follows:
\begin{codex}
?- gold_test(6,lprove,T,R).
false. % indicates that no false positive or negative is found

?- gold_test(6,badProve,T,R).
T =  (0->1->0->0->0->0->0),
R = wrong_failure ;
...
?- gold_test(6,badProve,T,wrong_success).
T =  (0->1->0->0->0->0->2) ;
...
\end{codex}

A more interesting case is when a prover is only guilty of false positives.
For instance, let's naively implement the intuition that
a goal is provable w.r.t. an environment {\tt Vs} if all its
premises are provable, with implication introduction assuming
 premises and success achieved when the environment is
 reduced to empty.
 
 \begin{code}
badSolve(A):-badSolve(A,[]).

badSolve(A,Vs):-atomic(A),!,memberchk(A,Vs).
badSolve((A->B),Vs):-badSolve(B,[A|Vs]).
badSolve(_,Vs):-badReduce(Vs).

badReduce([]):-!.
badReduce(Vs):-select(V,Vs,NewVs),badSolve(V,NewVs),badReduce(NewVs).
\end{code}

As the following test shows, while no tautology is missed,
the false positives are properly caught.
\begin{codex}
?- gold_test(6,badSolve,T,wrong_failure).
false.

?- gold_test(6,badSolve,T,wrong_success).
T =  (0->0->0->0->0->0->1) ;
...
\end{codex}

\subsection{Testing With Large Random Terms}

Testing  for false positives and false negatives for random terms proceeds in a similar
manner to exhaustive testing with terms of a given size.

Assuming Roy Dyckhoff's prover as a gold standard, we can find out that our
{\tt bprove/1} program can handle 20 terms of size 50
as well as the gold standard.
\begin{codex}
?- gold_ran_imp_test(20,100,bprove, Culprit, Unexpected).
false. % indicates no differences with the gold standard
\end{codex}

In fact, the size of the random terms handled by {\tt bprove/1}
makes using provers an appealing alternative to random lambda term
generators in search for very large  (lambda term, simple type) pairs.
Interestingly, on the side of random simply typed  terms, limitations come from  
their vanishing density, while on the other side
they come from the known PSPACE-complete complexity of the proof procedures.

\subsection{Scalability Tests}

Besides the correctness and completeness test sets described so far, one might want also
ensure that the performance of the derived provers scales up to larger terms. 
We  show here a few such performance tests and refer the reader to our benchmarks at:
{\small \url{https://github.com/ptarau/TypesAndProofs/blob/master/bm.pro}}.

Time is measured in seconds. The tables in Fig. \ref{tab} compare several provers on exhaustive "all-terms" benchmarks,
derived from our correctness test. 

\noindent
\begin{figure}
\begin{tabular}{||r||r|r|r||r||}
\cline{1-5}
\cline{1-5}
\multicolumn{1}{||c||}{{Prover}} & 
\multicolumn{1}{c|}{{ Size}} &
\multicolumn{1}{c|}{{ Positive}} &
\multicolumn{1}{c|}{{ Mix }} &
\multicolumn{1}{c|}{{ Total Time }}\\
\cline{1-5} 
\cline{1-5}
lprove & 13& 0.979 & 0.261 & 1.24 \\ \cline{1-5}
lprove & 14 & 4.551 & 5.564 & 10.116 \\ \cline{1-5}
lprove & 15 & 30.014 & 5.568 & 35.583 \\ \cline{1-5}
lprove & 16 & 3053.202 & 168.074 & 3221.277 \\ \cline{1-5}
\hline\hline
bprove & 13 & 0.943 & 0.203 & 1.147 \\ \cline{1-5}
bprove & 14 & 4.461 & 4.294 & 8.755 \\ \cline{1-5}
bprove & 15 & 32.206 & 4.306 & 36.513 \\ \cline{1-5}
bprove & 16 & 3484.203 & 129.91 & 3614.114 \\ \cline{1-5}
\hline\hline
dprove & 13 & 5.299 & 0.798 & 6.098 \\ \cline{1-5}
dprove & 14 & 23.161 & 13.514 & 36.675 \\ \cline{1-5}
dprove & 15 & 107.264 & 13.645 & 120.909 \\ \cline{1-5}
dprove & 16 & 1270.586 & 240.301 & 1510.887 \\ \cline{1-5}
\end{tabular}
\quad
\begin{tabular}{||r||r|r|r||r||}
\cline{1-5}
\cline{1-5}
\multicolumn{1}{||c||}{{Prover}} & 
\multicolumn{1}{c|}{{ Size}} &
\multicolumn{1}{c|}{{ Positive}} &
\multicolumn{1}{c|}{{ Mix }} &
\multicolumn{1}{c|}{{ Total Time }}\\
\cline{1-5} 
\cline{1-5}
hprove & 13 & 1.007 & 0.111 & 1.119 \\ \cline{1-5}
hprove & 14 & 4.413 & 1.818 & 6.231 \\ \cline{1-5}
hprove & 15 & 20.09 & 1.836 & 21.927 \\ \cline{1-5}
{\bf hprove} & {\bf 16} & {\bf 90.595} & {\bf 30.713} & {\bf 121.308} \\ \cline{1-5}
\hline\hline
eprove & 13 & 1.07 & 0.132 & 1.203 \\ \cline{1-5}
eprove & 14 & 4.746 & 2.27 & 7.017 \\ \cline{1-5}
eprove & 15 & 21.562 & 2.248 & 23.81 \\ \cline{1-5}
eprove & 16 & 97.811 & 43.18 & 140.991 \\ \cline{1-5}
\hline\hline
sprove & 13 & 1.757 & 0.173 & 1.931 \\ \cline{1-5}
sprove & 14 & 8.037 & 2.966 & 11.003 \\ \cline{1-5}
sprove & 15 & 38.266 & 2.941 & 41.208 \\ \cline{1-5}
sprove & 16 & 188.317 & 54.802 & 243.12 \\ \cline{1-5}
\end{tabular} 
\caption{Performance of provers on exhaustive tests (faster ones in the right table)
\label{tab}}
\end{figure}
First, we run them on the types inferred on all 
simply typed lambda terms of a given size. Note that some of the resulting types in this case can be larger and some smaller than the sizes of their inhabitants. We place them in the column {\em Positive} - as they are
known to be all provable.

Next, we run them on all implicational formulas of
a given size, set to be about half of the former (integer part of size divided by 2), as the number of these grows 
much faster. We place them  in the column {\em Mix} as they are a mix of provable and unprovable formulas.

The predicate  {\tt hprove/1} turns out to be an overall winner, followed closely by
{\tt eprove/1} that applies to implicational forms a technique borrowed from {\tt hprove/1} to quickly filter out
failing search branches.

Testing exhaustively on small formulas, while an accurate indicator for average speed, might not favor provers using more complex heuristics or extensive preprocessing, as it is the case of Dyckhoff's original {\tt dprove/1}.
 
We conclude that early rejection via the test we have discovered
in the nested Horn clause form is a clear separator between the slow provers in the left table
and the fast ones in the right table, a simple and useful ``mutation'' worth propagating to
full propositional and first order provers.

As the focus of this paper was to develop a  testing methodology for propositional theorem provers, we have not applied more intricate heuristics to further improve performance or to perform better on ``human-made'' benchmarks or compare
them on such tests with other provers, as there are no purely implicational tests among
 at the ILTP library \cite{iltp} at \url{http://www.iltp.de/}. On the other hand, for our full intuitionistic propositional provers at \url{https://github.com/ptarau/TypesAndProofs}, as well as our Python-based ones at \url{https://github.com/ptarau/PythonProvers}, we have adapted the  ILTP benchmarks on which we plan to report in a future paper.

\section{A Use Case: Finding Bugs in Theorem Provers}\label{bugs}

Transformations that result in equiprovable formulas can be used to find bugs in theorem provers that escape all human-made ILTP tests, as well as our own exhaustive test on  formulas of small size.

\subsection{Catching Bugs by Hardening Implicational Formulas Known-to-be Tautologies with the Mints Transformation}

We start with known tautologies in implicational fragment of intuitionistic propositional calculus obtained via the Curry-Howard correspondence as well as formulas in Full intuitionistic propositional calculus proven or disproven (by the same prover or other known to be correct prover), before applying the transformation.

Consequently, we obtain  harder to prove, significantly larger formulas, for which we know that they have originated from a smaller formula with known status as provable or unprovable.

As an example, we tested the fcube 4.1 intuitionistic propositional calculus prover \cite{fcube} available at \url{http://www2.disco.unimib.it/fiorino/fcube.html}. It is a very nice Prolog-based prover that, in our tests, has outperformed everything else on the ILTP human-made tests. It has also passed all our tests on formulas up to size {\bf 12}.

But testing against the Mints transform finds incompleteness bugs:
\begin{codex}
?- small_taut_bug(4,fcube).
unexpected_failure_on
0->1->2->3->0
<=>
(nv1->0->nv2)->(nv2->1->nv3)->(nv3->2->nv4)->(nv4->3->0)->
    ((0->nv2)->nv1)->((1->nv3)->nv2)->((2->nv4)->nv3)->
    ((3->0)->nv4)->nv1
\end{codex}
Fortunately, they seem to be fixed in the next version of {\bf fCube} available from the same site.

\subsection{Catching Bugs Using more General Intuitionistic Propositional Calculus Formulas}

We can catch a bug if the ``suspect''  disagrees with itself on the small easy formula and its hard transformed formula.

When acting on the transformed formula, with the original seen as an oracle, a prover can be found out as unsound if it proves a non-tautology and incomplete if it fails to prove a tautology.

Thus, we can use agreement with a trusted prover running on the small formula:
\begin{code}
mints_fcube(A):-mints(A,MA),fcube(MA).
\end{code}

\begin{codex}
?- gold_eq_neg_test(5,mints_fcube,Culprit,Unexpected).
Culprit = ~ (0<->(1<-> ~ (1<->0))), Unexpected = wrong_failure ;
Culprit = ~ (0<->(1<-> ~ (0<->1))), Unexpected = wrong_failure ;
...
\end{codex}
Note that 
{\tt gold\_eq\_neg\_test} compares behavior of a given prover against a trusted ``gold standard'' prover.
It is more relevant for human eyes to only display the source, before the transformation is applied.
In this case we find  formulas containing negation and equivalence on which
the prover obtained by applying of the Mints transform to the suspect fails the test, 
revealing the same incompleteness bug.

\section{Related Work}\label{rel}

The related work derived from Gentzen's {\bf LJ} calculus is in the hundreds if not in the thousands of papers and books. Space constraints limit our discussion to the most closely related papers,
directly focusing on algorithms for  implicational intuitionistic propositional logic, which, as decision procedures, ensure termination without a loop-checking mechanism.

Among them the closest are \cite{dy1,dy2}, that we have used as starting points
for deriving our provers. We have chosen  to implement the {\bf LJT} calculus directly
rather than deriving our programs from Roy Dyckhoff's Prolog code.
At the same time, as in Roy Dyckhoff's original prover, we have benefitted
from the elegant, loop-avoiding rewriting step also present
in Hudelmaier's work \cite{hud88,hud93} and originally due to Vorobiev \cite{vor58}.
Similar calculi, key ideas of which  made it into the Coq  proof assistant's code, are
described in \cite{HerIJT}.

On the other side of the Curry-Howard isomorphism, the thesis \cite{ben79}, described in full detail
in \cite{hindleyTypes}, finds and/or counts inhabitants of simple types in long normal form.
But interestingly, these algorithms have not crossed, at our best knowledge,
to the other side of the Curry-Howard isomorphism, in the form of theorem provers.

Using hypothetical implications in Prolog, although all with a different semantics
than Gentzen's {\bf LJ} calculus or its {\bf LJT} variant, 
go back as early as \cite{nprolog84,nprolog85}, followed
by a series of $\lambda$Prolog-related publications, e.g., \cite{miller12}. 
The similarity to the propositional subsets of
N-Prolog \cite{nprolog85} and $\lambda$-Prolog \cite{miller12} comes from
their close connection to intuitionistic logic. The hereditary Harrop formulas 
of  \cite{miller12}, when restricted to their implicational subset, are more easily
computable with a direct mapping to Prolog, without the need of theorem prover.
While closer to an {\bf LJ}-based
calculus, the execution algorithm of \cite{nprolog85} 
uses restarts on loop detection instead of ensuring termination 
 along the lines the {\bf LJT} calculus.
In \cite{TDF:asian96} 
backtrackable linear and intuitionistic assumptions that mimic
the implication introduction rule are used, but they do not involve
arbitrarily deep nested implicational formulas.

Overviews of closely related calculi, using the implicational subset
of propositional intuitionistic logic are \cite{gabbay2002goal,dy2}.

For generators of random lambda terms and related functional programming constructs
we refer to \cite{qcheck,palka11}. 
We have shared with them the goal of achieving high-probability correctness via automated combinatorial testing. 
Given our  specific focus on propositional provers, we have  been able to use all-term and all-formula generators as well as comparison mechanisms with "gold-standard" provers.
We have also taken advantage of the Curry-Howard isomorphism between types and formulas to provide an initial set of known tautologies, usable as "bootstrapping mechanism" allowing to test our provers independently from using a "gold-standard".

Generators for closed simply-typed 
lambda terms, as well as their normal forms,
expressed as functional programming algorithms,
are given in \cite{grygielGen}, derived from
combinatorial recurrences for closed terms
and additional filtering for typability.

The idea of using Boltzmann samplers for generating random lambda terms
was first introduced in \cite{binlamb}.
Random lambda term generation with focus on 
practical uses in testing programming languages and proof assistants
is covered in \cite{palka11},  which reports
using them to find  bugs in the GHC Haskell compiler.

We have used extensively Prolog as a meta-language 
for the study of combinatorial and computational properties of lambda 
terms in papers like \cite{padl17,ppdp15tarau} covering different families
of terms and properties.

The idea to use types inferred for lambda terms as formulas 
for testing theorem provers 
originates in \cite{padl19}.  
The current paper extends this line of research to the full
intuitionistic propositional logic, provides a family of algorithms
for exhaustive and  random tautology generators (including
the combinator-based generator of random tautologies).
It also describes implementation of a rich set of formula transformers,
among which, the one from disjunction-free formulas to Nested Horn Clauses.
This, together with the   $O(n~log(n))$-space Nested Horn Clause prover
covers the highly expressive N-Prolog 
subset of propositional intuitionistic logic \cite{nprolog84}.


\section{Conclusions and Future Work}\label{conc}

Correctness and scalability testing of theorem provers is likely to impact on their application to formal methods and proof assistants.
 Besides the ability to  also evaluate  scalability and performance of theorem provers, components of our combinatorial generation library, released
as open source software,
 have good chances to be reused as a testing harness for theorem provers for intuitionistic, temporal, modal logic, as well as
SAT, ASP or SMT solvers, with structurally similar formulas. 
Generators for typed lambda terms can also be reused in testing type inference algorithms
in newly implemented programming languages or in wrappers adding type systems for languages like Python and Javascript. Large simply typed lambda terms can be used for performance and scalability tests for run-time systems of functional language implementations.

Future work will focus on extending our formula generation techniques to test provers for intuitionistic first order logic and some of its weaker sub-logics.

\section*{Acknowledgement} 
We thank the participants to the CLA'2018 and CLA'2019 for their comments and suggestions.

\bibliographystyle{unsrt}
\bibliography{tarau,theory,proglang,biblio}

\end{document}